\documentclass[conference]{IEEEconf}

\usepackage{graphicx}
\usepackage{multirow}
\usepackage{titlesec}
\usepackage{balance} 
\usepackage{stfloats}
\usepackage{xcolor}

\usepackage{cite}
\usepackage{amsmath}
\usepackage{amssymb}
\usepackage{textcomp}
\usepackage{xspace}

\usepackage{listings}
\usepackage{url}
\usepackage[export]{adjustbox}
\usepackage[caption=false]{subfig}
\usepackage[linesnumbered,ruled,vlined]{algorithm2e}
\usepackage{mathtools}
\usepackage{wrapfig}
\usepackage{booktabs}
\usepackage{here}
\usepackage{mdframed}
\usepackage{rotating}

\usepackage{article}
\usepackage{xrtable}

\renewcommand\thesection{\arabic{section}} 

\renewcommand\thesubsectiondis{\thesection.\arabic{subsection}}

\renewcommand\thesubsubsectiondis{\thesubsectiondis.\arabic{subsubsection}}

\newtheorem{definition}{Definition}

\mdfsetup{ %
    skipabove=0em, skipbelow=.5em, %
    innertopmargin=.25em, innerbottommargin=.25em, %
    innerleftmargin=.25em, innerrightmargin=.25em, %
    backgroundcolor=brown!10, %
    linecolor=black %
}
\mdfdefinestyle{answer}{ %
    backgroundcolor=green!10, %
    roundcorner=10pt
}

\begin{document}

\title{\textbf{\Large Bit-Precise Conformance Testing of Simulink Model Checkers\\}}

\author{Daisuke Ishii$^{1,*}$, Takashi Tomita$^{1}$, Toshiaki Aoki$^{1}$, and Hideaki Takai$^{2}$\\
	\normalsize $^{1}$Japan Advanced Institute of Advanced Science, Ishikawa, Japan\\
	\normalsize $^{2}$GAIO Technology Co. Ltd., Tokyo, Japan\\
	\normalsize \{dsksh,tomita,toshiaki\}@jaist.ac.jp, takai.h@gaio.co.jp\\
	\normalsize *corresponding author
}


\maketitle
\begin{abstract}
MATLAB/Simulink provides a practical modeling language and a simulation engine for the development of cyber-physical systems.
To ensure the quality of the developed models, there are formal verification tools available, such as Simulink Design Verifier (\SLDV) and third-party SMT-based model checkers (\SmtMC).
However, due to the absence of a semantics of Simulink that covers every element of models and the details of its numerical behavior, the reliability of the model checkers themselves is often doubtful, as they may analyze models differently from the simulator.
This work aims to verify the quality of the Simulink model checkers by addressing the following items.
1)~Formalization of the basic block types of Simulink.
It involves defining block type feature sets and the bit-precise behavior of the blocks.
2)~A method for testing bit-precise conformance relations among the tools for each block type. The pass rate of our test suite measures (i)~conformance of model checking results with simulation results by Simulink and (ii)~conformance between the results of \SmtMC and \SLDV.
3)~Experiments on 10~block types. 
We confirmed that \SmtMC efficiently passed all test cases, while \SLDV achieved pass rates of only $94$--$96\%$ and $80$--$90\%$ for conformance~(i) and (ii), respectively.
We analyzed the causes of failed tests, such as errors, corner cases, and timeouts.
\end{abstract}
\IEEEoverridecommandlockouts
\vspace{1.5ex}

%
\IEEEpeerreviewmaketitle

\section{Introduction}

An efficient development method for cyber-physical systems (CPSs)~\cite{leeIntroduction2017} is to mimic real-world systems using models constructed from blocks and deepen understanding through simulation and analysis.
\emph{MATLAB/Simulink} (Sect.~\ref{s:simulink}) provides a language for describing models as block diagrams and a toolchain including a numerical simulator and additional toolboxes.

For the quality assurance of safety-critical \Simulink models, 
as demanded by standards such as ISO~26262 for automotive and DO-178C for the aircraft domain~\cite{ibrahimState2021}, 
the \emph{Simulink Design Verifier} toolbox (\SLDV) provides a \emph{model checking (MC)} function (Sect.~\ref{s:simulink:mc}).
Namely, it analyzes the contents of models in a formal manner and determines whether certain properties are satisfied.
Third-party tools are also available to supplement the functions and performance of the standard tools.
For example, numerous MC methods and tools using \emph{SMT solvers} have been proposed, e.g., \cite{schrammelIncremental2017,bourbouhCoCoSim2020,ishiiSMTBased2022,cleavelandTwo2025}.

An important detail in \Simulink MC is analyzing numerically interpreted models at bit-level precision (Sect.~\ref{s:simulink:prod}).
Although models are mathematically hybrid systems of continuous and discrete behaviors, \Simulink simulates them based on machine-representable numbers (e.g. \inlst{double} and \inlst{int8}) and discrete steps.
\SLDV interprets models in the same way, meaning that the results of MC are expected to match those of the simulation.
However, third-party methods often interpret them as idealized mathematical models.
While many formalization and MC methods have been proposed, research considering bit-precise behavior is scarce (Sect.~\ref{s:related:prec}).

Although the quality of the tools is important (Sect.~\ref{s:related:qual}), 
\Simulink and its toolchain have issues that come with their proprietary nature.
First, the documentation and semantics of the \Simulink language are available (Sect.~\ref{s:related:formal}), but the corner cases and bit-precise semantics are often unclear.
\Simulink blocks have many parameters for each block type, which results in a wide variety of instances;
it is difficult to comprehensively describe how each instance operates in each step.
Second, there is room for improvement in the reliability of official tools (\SLDV is the target of our test).
As stated in the release notes, bugs that may lead to incorrect results have been fixed in almost every version.
When we try using \SLDV, it is not uncommon to encounter errors or ``undecided'' outputs.
In addition, \SLDV has scalability issues in analyzing large models. 
Finally, when third-party tools are used in development (we also target an SMT-based tool), it becomes difficult to check and qualify the conformance of the tools to the semantics and the behavior of \Simulink models.
Although most methods have been verified for correctness through their formalization in technical documents and through testing of their implementations, users still commonly check the detailed conformance by comparing the results with the \Simulink simulator.

In this work, we aim to formalize the \Simulink language and clarify the conformance degree 
of the \Simulink MC tools, i.e., \SLDV and an \emph{SMT-based model checker} (\SmtMC)~\cite{ishiiSMTBased2022}.
In particular, we examine the underlying numerical computations at the bit level and carry out precise formalization and testing.
Our contributions are summarized as follows.

\vspace{.5em}
\emph{1)~Formalization of basic \Simulink block types}.
We specify a subset of the \Simulink language through two efforts:
First, we extract \emph{block type feature models} (Sect.~\ref{s:method:bt})
from the official documentation for 10~block types.
Second, we describe the bit-precise operational semantics of blocks 
as an extended \emph{\SmtMC encoder} (Sect.~\ref{s:simulink:mc})
that translates blocks into predicate logic formulas. 
It replicates \Simulink's interpretation by describing the behavior of each block with logic formulas involving machine-representable integers and real numbers, represented by bit vectors.

\begin{figure}[!t] \centering
    \includegraphics[width=.4\textwidth]{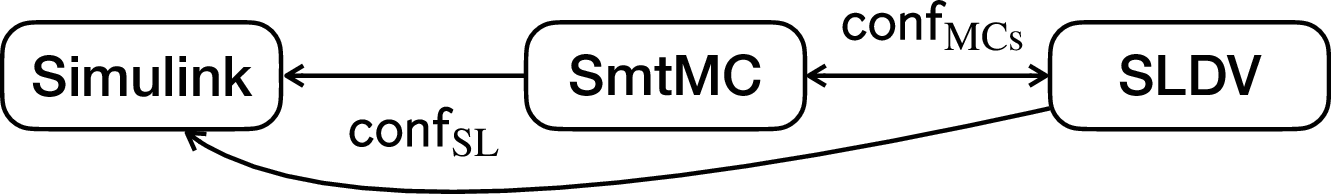} 
    \caption{Tested conformance relations among tools.}
    \label{f:conf}
\end{figure}

\vspace{.5em}
\emph{2)~A bit-precise testing method for checking the conformance among \Simulink, \SmtMC and \SLDV} (Sect.~\ref{s:method}).
We define $\ConfSL$ to ensure that \SmtMC and \SLDV interpret \Simulink models in exactly the same way as the \Simulink simulator does, and $\ConfMCs$ to ensure the conformance of the MC functionality, as shown in Fig.~\ref{f:conf}.
To verify $\ConfSL$ and $\ConfMCs$,
we propose a combinatorial testing 
method using the \Simulink simulator as an oracle to make bit-precise comparison of the results (we also use \SmtMC as an oracle for $\ConfMCs$).
It assumes a subset of block type instances and a subset of input signals,
generates test cases (TCs) covering the subsets in a pairwise combinatorial manner, and checks that each TC does not violate the conformance;
as a result, it outputs the percentage of passed TCs in the test suite.
Our method systematically tests the correctness and performance of model-based development tools based on detailed comparison among them.
We are not aware of many other examples of tool evaluation that can inspect this level of precision.

\vspace{.5em}
\emph{3)~Experimental results} (Sect.~\ref{s:xp} and Sect.~\ref{s:discuss}).
Our experimental results provided the tools with new quantitative evaluation metrics, i.e., the degree of bit-precise conformance (RQ1). 
Regarding $\ConfSL$, we confirmed that \SmtMC passed all TCs;
\SLDV failed to achieve an all-pass result, with average pass rates of $94\%$ and $96\%$ for versions R2020b and R2025b, respectively.
Regarding $\ConfMCs$, the average pass rates when comparing \SmtMC with \SLDV were $80\%$ and $90\%$ for R2020b and R2025b, respectively.
Notably, while \SmtMC was able to process all TCs efficiently on average, a few percent of \SLDV tests resulted in timeouts or errors.
Accordingly, we analyzed the results to clarify the criteria for model instances supported by the tools (RQ2), and
identified sets of TCs that produce incorrect results, TCs that cause errors, and TCs that take a long time;
this thorough analysis helped us fix many bugs in \SmtMC.
In addition, we compared the execution times when processing concrete TCs and abstract TCs with \SmtMC and \SLDV (RQ3);
as a result, \SmtMC was fast in most cases, but \SLDV was faster in cases where the \inlst{Product} block contained uncertainty.

%

\section{Simulink}
\label{s:simulink}

\Simulink\footnote{\url{https://www.mathworks.com/products/simulink.html}}
is a popular MATLAB toolbox for the model-based design and prototyping of CPSs.
It provides a graphical modeling language to describe dynamical systems that consist of interacting blocks and subsystems, and a numerical simulator that computes model behaviors on discrete steps, which correspond to either fixed or variable sample time.

\emph{Simulink models} are diagrams structured as hierarchical directed graphs with \emph{lines} (edges) and \emph{blocks} (vertices).
Each component of models, i.e., blocks and user-defined models (subsystems), 
represents a function that maps between sets of input and output signals, and 
behaves as a reactive system interacting in a synchronous manner.

\begin{figure}[!t] \centering
    \subfloat[\label{f:ex:model} A model diagram.]{%
        \includegraphics[width=.475\textwidth]{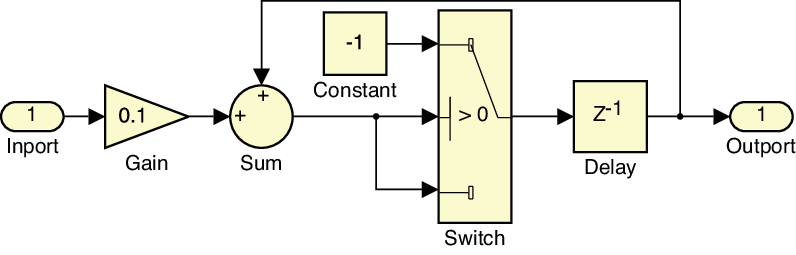} 
    } 
    \vspace*{.5em}
    \subfloat[\label{f:ex:exec} An execution.]{%
        \includegraphics[width=.4\textwidth]{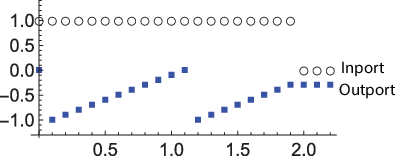}%
    }
    \caption{Example \Simulink model.}
    \label{f:ex}
\end{figure}

An example \Simulink model is shown in Fig.~\ref{f:ex:model}, which
describes an accumulator for input values that may be reset if a condition is met.
The counter is described by a feedback loop that consists of block instances of types \inlst{Inport}, \inlst{Gain}, \inlst{Sum}, \inlst{Constant}, \inlst{Switch}, \inlst{Delay}, and \inlst{Outport}.
Each block type has \emph{parameters} such as 
\verb|Value| setting the output value~$-1$ of \inlst{Constant},
\verb|Inputs| setting the number of arguments~$2$ of \inlst{Sum}, and
\verb|InitialCondition| setting the initial output value~$0$ of \inlst{Delay}.

We can interpret \Simulink models as transition systems. 
\begin{definition} \label{d:ts}
    Let $\VIn$, $\VOut$ and $\VSt$ be \emph{input}, \emph{output}, and \emph{state variables}.
    For a (vector-valued) variable $V=(v_1,\ldots,v_n)$, we denote its \emph{domain} $D_1\!\times\!\cdots\!\times\!D_n$ by $D(V)$.
    A \emph{transition system} $(\VIn,\VOut,\VSt,\Init,\Trans)$ consists of an \emph{initial condition} $\Init \subseteq D(\VSt)$ and a \emph{transition relation} $\Trans \subseteq D(\VSt)\!\times\!D(\VSt)\!\times\!D(\VIn)\!\times\!D(\VOut)$.
\end{definition}

Variables are typed as \verb|double| (double-precision floating-point (FP) numbers), \verb|int32| (32-bit signed integers), \verb|string|, etc.
Not only do they represent scalar values, they can also have a composite type such as vectors of $n$ values, $m \times n$ matrices, higher-dimensional arrays and buses. 
For example, the model in Fig.~\ref{f:ex} is interpreted with variables $i_1$ and $o_1$,
representing the input and output values of the \inlst{Inport} and \inlst{Outport} blocks, and
$s_1$, representing the next output values of \inlst{Delay}.
Let $e$ be the expression $0.1 i_1 + s_1$. 
Then, the initial condition is $\Init :\leftrightarrow (s_1 = 0)$ and 
the transition relation is $\Trans :\leftrightarrow ((o_1 = \mathrm{if}\ e > 0\ \mathrm{then}\ {-1}\ \mathrm{else}\ e) \land s_1' = o_1)$ where $s_1'$ represents the next state.
Each variable are typed as \verb|double|, etc. (it can also be typed as a composite).
We assume $M$ is deterministic, i.e. $\Trans(s_1,i_1,o_1,s_1')$ represents a map from any value $(s_1,i_1)$ to a single value $(o_1,s_1')$.

For a variable $V$, we call a finite sequence $\beta(0) \cdots \beta(k-1)$ of $k$ values in the domain $D(V)$ a \emph{signal} and denote it by $\beta^{[k]} \in D(V)^k$.
Numerical simulation using the \Simulink tool computes the output signals of models, given input signals.
They are formalized as executions of transition systems.
\begin{definition} \label{d:exec}
    Assume a model $M = (\VIn,\VOut,\VSt,\Init,\Trans)$ and an input signal $\iota^{[k]} \in \Dom{\VIn}^k$.
    Let $\sigma^{[k+1]}$ and $\Omic^{[k]}$ be signals in $\Dom{\VSt}^{k+1}$ and $\Dom{\VOut}^k$, respectively.
    Then, an \emph{execution} is
    \begin{equation*}
        \sigma({0}) \xrightarrow{\iota(0)/\Omic(0)} \sigma(1) 
        ~\cdots~ \sigma({k\!-\!1}) \xrightarrow{\iota({k-1})/\Omic({k-1})} \sigma({k}),
    \end{equation*}
    where 
    $\Init(\sigma({0}))$ holds and $\Trans(\sigma({j}), \AB \iota(j), \AB \Omic(j),\AB \sigma({j+1}))$ holds for every $j \in \{0,\ldots, k\!-\!1\}$.
    We denote the \emph{output signal} $\Omic^{[k]}$ by $\Int{M}(\iota^{[k]})$.
\end{definition}


Input and output signals of the example model are shown in Fig.~\ref{f:ex:exec}.
When actually simulating with \Simulink, we need to specify the input signal using a harness outside the description of Fig.~\ref{f:ex:model}.

In this work, we assume a fixed sample time.
The transitions described above are based on discrete steps, each of which corresponds to the step time, and signals can be regarded as functions of time;
in Fig.~\ref{f:ex:exec}, the step time is set to $0.1$.

\subsection{\inlst{Product} Block Type and its Bit-Precise Semantics}
\label{s:simulink:prod}

\begin{figure}[!t] \centering
    \subfloat[\label{f:ex:prod1}``\texttt{2}'' or ``\texttt{**}'']{%
        \includegraphics[width=.125\textwidth,valign=t]{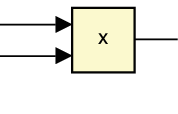}%
    } \qquad
    \subfloat[\label{f:ex:prod2} ``\texttt{//}'']{%
        \includegraphics[width=.125\textwidth,valign=t]{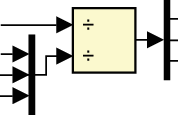}%
    } \qquad
    \subfloat[\label{f:ex:prod3} ``\texttt{1}'' or ``\texttt{*}'']{%
        \includegraphics[width=.125\textwidth,valign=t]{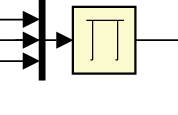}%
    }
    \caption{Instances of \inlst{Product} corresponding to the values of \texttt{Inputs}.}
    \label{f:ex:prod}
\end{figure}

This work addresses the detailed semantics of basic block types. 
Here, we explain one of them, \inlst{Product}.
Its block instance has one or more variable input ports $i_1$,\ldots, $i_n$ and one output port $o_1$, and represents a multiplication of the input signal values.
Depending on the values of the \verb|Inputs| parameter, an instance takes various shapes as shown in Fig.~\ref{f:ex:prod}.
The operation changes depending on whether the inputs are composite (e.g. vector) signals or not
(it also depends on parameters \verb|Multiplication| and \verb|CollapseMode|, but we assume default settings, i.e., \inlst{Element-Wise} and \inlst{All dimensions}).
The instances in the figure represent
$o_1 = i_1 \times i_2$ (Fig.~\ref{f:ex:prod1}),
$o_{1,j} = (1 \div i_1) \div i_{2,j}$ for $1 \le j \le 3$ (Fig.~\ref{f:ex:prod2}), and
$o_1 = (i_{1,1} \times i_{1,2}) \times i_{1,3}$ (Fig.~\ref{f:ex:prod3}).

\begin{table}[!t]
    \centering
    \footnotesize
    \caption{\texttt{int8} operation examples of \texttt{Product}.} \label{t:prod}
    \setlength{\tabcolsep}{.5em}
    \begin{tabular}{@{}lrrrcc@{}}
        \toprule
        \verb|Inputs| & $i_1$ & $i_2$ & $o_1$ & \verb|Saturate| & \verb|RndMeth| \\
        \midrule
        ``\verb|2|''  & $127$ & $ 2$ & $ -2$ & false & -- \\
        ``\verb|2|''  & $127$ & $ 2$ & $127$ & true  & -- \\
        ``\verb|/*|'' & $  0$ & $ 2$ & $127$ & --    & -- \\
        ``\verb|/*|'' & $  3$ & $12$ & $  4$ & --    & -- \\
        ``\verb|//|'' & $  2$ & $ 2$ & $  0$ & --    & \verb|Floor| \\
        \bottomrule
    \end{tabular}
\end{table}

By default, \Simulink handles signal values as \verb|double| type, and 64-bit FP arithmetic is performed, following the IEEE-754 standard~\cite{ieee75420082008}. 
The result of each arithmetic operation is rounded to a representable FP number according to the specified mode.
Special values e.g. $+\infty$ and $\Nan$ are used to represent special operation results.
For example, when the result of multiplication exceeds the representable range, it becomes $+\infty$ with a rounding mode toward the positive direction.
Operations with other FP types, \verb|float| (32~bits) and \verb|half| (16~bits), also conform to the standard.

\Simulink handles (signed or unsigned) integers from 8~bits to 64~bits.
Integer arithmetic overflows can result in a variety of outcomes.
The behavior is controlled via setting the parameters, e.g., \verb|SaturateOnIntegerOverflow| (we abbreviate it as \verb|Saturate|) that takes a boolean value, and \verb|RndMeth| (rounding method) that takes one of seven modes.

Some tricky examples are shown in Tab.~\ref{t:prod}; the first column shows the values of the \texttt{Inputs} parameter; each row represents an \verb|int8| scalar operation with a setting of three parameters (`--' indicates that the values make no difference).
As shown in the first two cases, the value of \verb|Saturate| determines whether the results are wrapped or saturated.
In the third example, the first input is interpreted as a division by zero and it results in $127$, but in that case, the result of multiplication with $i_2$ is not wrapped; further operation results after division by zero are not wrapped.
In the fourth example, the first division number is applied to the second argument as $12/3$.
The operation is not immediately performed or rounded; instead, the operands are reordered to improve accuracy.
This is not the case for the last example; the operations in $(1/2)/2$ are performed with left associativity.

We suspect that sufficiently accurate real number representations (e.g. \verb|double| values for \verb|int8|) are used internally for integer operations; then, operations are performed simply by left associativity. 
The third example can be explained as the result of such an accurate operation;
the last example internally results in $0.25$ and is then rounded to $0$.

\subsection{\Simulink Model Checking}
\label{s:simulink:mc}

\SLDV\footnote{\url{https://www.mathworks.com/products/simulink-design-verifier.html}}
and tools based on SMT solvers such as \cite{schrammelIncremental2017,bourbouhCoCoSim2020,ishiiSMTBased2022,cleavelandTwo2025} provide functionality for formally analyzing \Simulink models;
we refer to such tools collectively as \emph{Simulink model checkers} or \emph{MC tools}.
They are given additional descriptions that specify \emph{properties} on the behavior (signals) of models. 
Many tools support the MC of safety properties.
Given a model $M$ and a property $\varphi$, a model checker returns whether $M$ satisfies $\varphi$, i.e., every (finite and infinite) trace of $M$ satisfies $\varphi$;
we denote this fact checked by a tool $\mathit{MC}$ by $M \models_\mathit{MC} \varphi$.

In our method, we consider only bounded safety properties, where both the satisfiability and unsatisfiability of a property can be confirmed by bounded-length executions.
Let $M = (I,O,S,\Init,\Trans)$ be a \Simulink model and $\varphi$ a property.
The properties we consider are in the form of $\Box(C(\VIn^{[k]}) \to C'(\VOut^{[k]}))$ where 
$\VIn^{[k]}$ and $\VOut^{[k]}$ are variables representing length-$k$ input and output signals, and 
$C$ and $C'$ are constraints that refer only to values at specific steps within the initial $k$ steps of the argument signal.
Obviously, we can check them by examining the executions with $k$ steps.
We are interested in verifying the basic functions of the MC tools, so we will limit ourselves to this bounded fragment.

\subsubsection{\SLDV}
The toolbox provides property blocks that can specify safety properties to annotate models.
Using its ``Property Proving'' function, users can analyze the annotated properties of a model and obtain reports on their validity.
The results are displayed in the GUI and output as HTML reports and binary files.
The underlying MC algorithm is not public, but it seems to involve multi-stage analysis and model approximation.
It is acknowledged by MathWorks that it is based on a model checking module,\footnote{\url{https://www.mathworks.com/help/sldv/ug/acknowledgments.html}} which appears to employ a formal proof method~\cite{sheeranstalmarck}.
\SLDV interprets models conforming to the \Simulink tool, and for numerical operations, it first appears to analyze using a rational approximation, and then more precisely.

\subsubsection{\SmtMC}
It is an SMT-based model checker provided as a supplemental functionality in PROMPT~V2~\cite{ishiiSMTBased2022}, a third-party tool for the automated testing of \Simulink models.\footnote{\url{https://www.en.gaio.co.jp/products/prompt-2/}}
It supports invariants (formulas on input and output variables enclosed within the $\Box$ operator) attached to the input \Simulink model.
\SmtMC encodes the model and the property into a predicate logic formula in the SMT-LIB format~\cite{barrettSMTLIB2025} (encoding examples are shown in Fig.~\ref{f:smt:exec} and in the Appendix).
For the behaviors involving FP numbers, we describe them based on the \inlst{FloatingPoint} theory of SMT-LIB.
For integers, we describe them based on the \inlst{FixedSizeBitVectors} theory, using user-defined arithmetic and rounding operators.
\SmtMC then checks its satisfiability using the \Impl{Z3} SMT solver.
The encoder of \SmtMC offers two options, one that approximates numerical values and one that handles them exactly based on bit vectors;
we use solely the latter in this work.
Furthermore, throughout our experiments, the encoder was gradually replaced with a modified version.


%

\section{Proposed Method}
\label{s:method}

\begin{figure}[t]
    \centering
    \includegraphics[width=.485\textwidth]{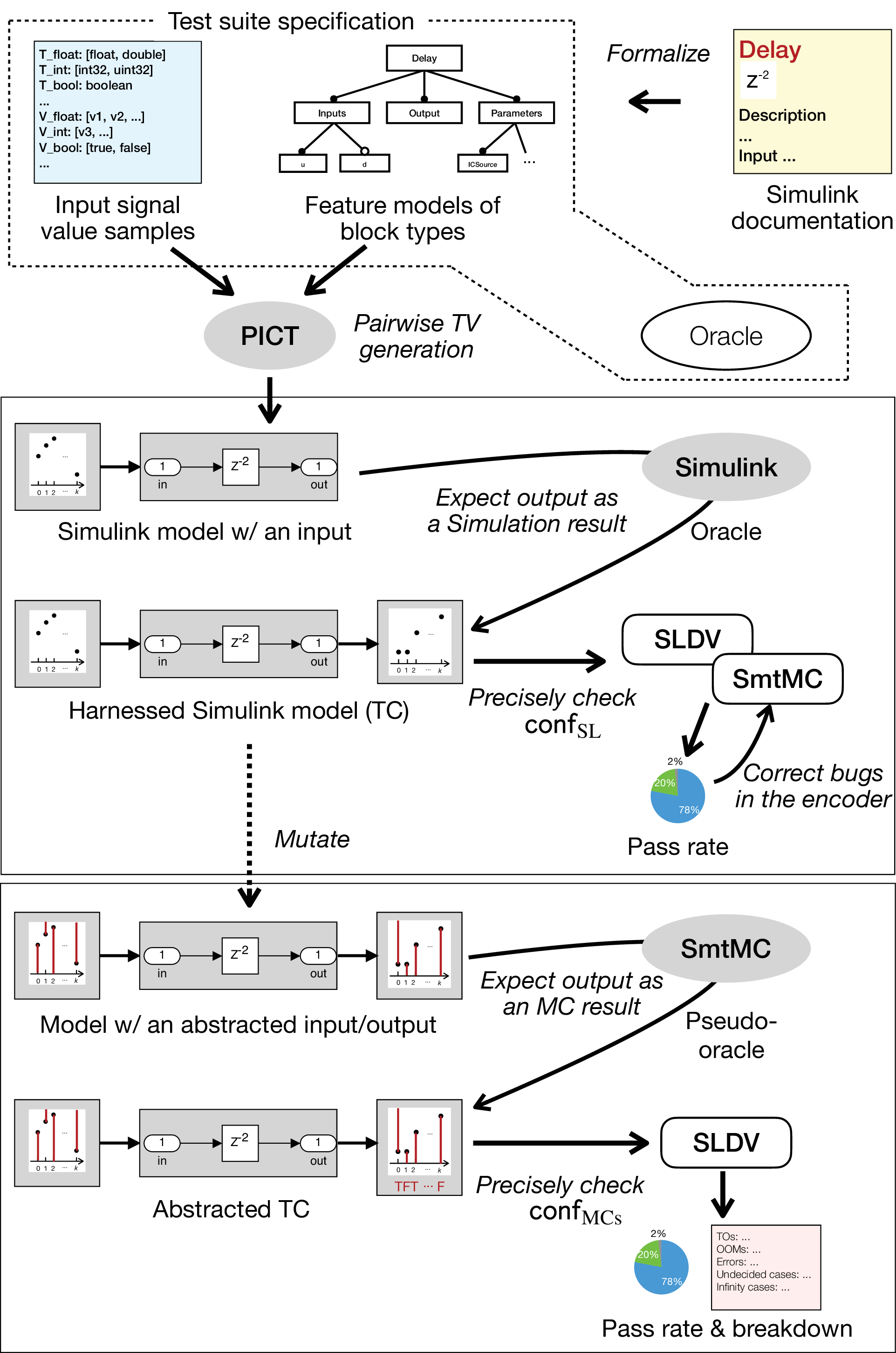} 
    \caption{Overview of the testing process.}
    \label{f:process}
\end{figure}

We aim at providing a means to check the conformance among \Simulink and MC tools.
We consider two kinds of conformance, $\ConfSL$ and $\ConfMCs$ 
(Sect.~\ref{s:method:conf}).
The relation $\ConfSL(\mathit{MC})$ expresses that the result of analyzing a \Simulink model using an MC tool $\mathit{MC}$ conforms to 
the behaviors of the model simulated with \Simulink (which is considered the oracle).
The relation $\ConfMCs(\mathit{MC}_1, \AB \mathit{MC}_2)$ represents whether the properties of a model checked by $\mathit{MC}_1$ can also be checked by $\mathit{MC}_2$ and vice versa; 
this allows us to compare the range of properties that MC tools can handle, bit-level precision, and performance.

Because the space of arbitrary input models and signals is enormous,
the proposed method checks the conformance partially with \emph{combinatorial testing}~\cite{czerwonkaPairwise2006,kuhnCombinatorial2015}, which measures the pass rate.
As a result, our method cannot guarantee the conformance, but it can detect non-conforming examples.
\begin{definition} \label{d:pr}
    Consider a {test suite specification} that provides a set $F$ of test factors, i.e. block features and input signal parameters.
    Let $S$ be a set of TCs that covers all pairwise combinations of factor levels in $F$, and $n$ be the number of TCs in $S$ that satisfy the condition~\eqref{eq:c1} or \eqref{eq:c2}.
    The \emph{pass rate} is the percentage of $n$ in $|S|$.
\end{definition}

The process illustrated in Fig.~\ref{f:process} checks $\ConfSL(\SmtMC)$, $\ConfSL(\SLDV)$ and $\ConfMCs(\SmtMC, \AB \SLDV)$.
Prior to testing, we prepare a \emph{test suite specification} (Sect.~\ref{s:method:spec}) that describes subsets of the \Simulink language and input signal values, in which the former subset is specified with a \emph{block type feature model} (Sect.~\ref{s:method:bt}).
%
We execute a combinatorial testing procedure (Sect.~\ref{s:method:proc})
in two ways for each of $\ConfSL$ and $\ConfMCs$;
it generates TCs in a pairwise combinatorial manner on the test suite factors, and
outputs a \emph{pass rate} (Def.~\ref{d:pr}).
The procedure consists of the following steps.
\begin{enumerate}
    \item A source of TCs is a \emph{test vector} generated in a combinatorial manner (we use \PICT~\cite{czerwonkaPairwise2006}) based on a test suite specification.
    We first prepare a \Simulink model that consists of a single target block instance and an input signal builder from a test vector.
    \item Next, we feed the model to an \emph{oracle}; depending on the conformance to be verified, we use \Simulink or \SmtMC as an oracle.
    Accordingly, we add a conformance verdict based on the output from the oracle to the TC model.
    %
    \item Once the TC is fed to the MC tool under test, the output (whether the MC result is not violating the conformance) is aggregated in the pass rate. 
    We also manually analyze all failure cases comprehensively.
    For the first test, we fix bugs in \SmtMC accordingly;
    in the second test, we classify the results and identify the causes.
\end{enumerate}

Sect.~\ref{s:impl} describes some details of the implementation.

\subsection{Conformance Relations}
\label{s:method:conf}

Following Def.~\ref{d:exec}, \Simulink interprets a model $M = (\VIn,\VOut,\VSt,\Init,\Trans)$ as a map $\Int{M}_\mathrm{SL} : D(\VIn)^k \to D(\VOut)^k$ from input signals to output signals.
As described in Sect.~\ref{s:simulink:mc}, MC tools (denoted by ``$\mathit{MC}$'') are fed $M$ and check whether a safety property $\varphi$ is satisfied (it is denoted by $M \models_\mathit{MC} \varphi$).
Accordingly, we consider two kinds of conformance relations among \Simulink and MC tools.
\begin{definition} \label{d:conf}
    The \emph{conformance} $\ConfSL(\mathit{MC})$ of an MC tool $\mathit{MC}$ to \Simulink is to satisfy 
    \begin{equation*} \label{eq:c1}
        M \models_\mathit{MC} \Box(\VIn^{[k]} = \iota^{[k]} ~\to~ \VOut^{[k]} = \Int{M}_\mathrm{SL}(\iota^{[k]}))
        \tag{$\ConfSL$}
    \end{equation*}
    for arbitrary models $M$ and input signals $\iota^{[k]}$ of arbitrary lengths $k$.
    $\VIn^{[k]}$ and $\VOut^{[k]}$ are variables representing input and output signal values from step $0$ to $k-1$.
    Let $\mathit{MC}_1$ and $\mathit{MC}_2$ be MC tools.
    The \emph{conformance} $\ConfMCs(\mathit{MC}_1,\mathit{MC}_2)$ between $\mathit{MC}_1$ and $\mathit{MC}_2$ is to satisfy
    \begin{equation*} \label{eq:c2}
        M \models_{\mathit{MC}_1} \varphi ~~\leftrightarrow~~
        M \models_{\mathit{MC}_2} \varphi
        \tag{$\ConfMCs$}
    \end{equation*}
    for arbitrary models $M$ and arbitrary safety properties $\varphi$.
\end{definition}

Reliable conformance checking will be possible if bit-precise MC tools and bit-precise comparisons of signal values in the condition~\eqref{eq:c1} are assumed.

Many existing conformance relations~\cite{tretmansFormal1992,broyModelBased2005,hieronsUsing2009,aertsModelBased2017} are defined between models $M_1$ and $M_2$ (i.e. state transition systems) based on the sets $\mathcal{L}(M_1)$ and $\mathcal{L}(M_2)$ of their traces (e.g. input and output signals).
In $\ConfSL$ and $\ConfMCs$, 
we compare the results of tools $T_1$ and $T_2$, each of which analyzes models $M$ by interpreting the trace sets $\mathcal{L}_{T_1}(M)$ or $\mathcal{L}_{T_2}(M)$, respectively.

\subsection{Block Type Feature Models}
\label{s:method:bt}

\begin{figure}[t]
\centering
\begin{lstlisting}[basicstyle=\ttfamily\footnotesize,frame=single]
BlockType: Product2
BlockPath: 'simulink/Math Operations/Product'
Inputs:
- { Name: 'Port_1', 
    Multiplicity: ['scalar', 'vector'], 
    Domain: [T_float, T_int, T_bool] }
- { Name: 'Port_2', 
    Multiplicity: ['scalar', 'vector'], 
    Domain: '<Port_1>' }
Output:
{ Multiplicity: ['scalar', 'vector'], 
  Domain: '<Port_1>',
  Constraint: 'IF (<im(Port_1)> = "vector" 
      OR <im(Port_2)> = "vector") 
    THEN <om(Output)> = "vector" 
    ELSE <om(Output)> = "scalar"' }
Parameters:
- { Name: 'Inputs', 
    Domain: ['2', '*/', '/*', '//'] }
- { Name: 'SaturateOnIntegerOverflow', 
    Domain: ['off', 'on'] }
- { Name: 'RndMeth', 
    Domain: ['Ceiling','Nearest',...] 
      # Some values are omitted. }
# Other parameters are omitted.
\end{lstlisting}
    \caption{Example block type specification.}
    \label{f:bspec}
\end{figure}

Each \Simulink block type is specified with input and output interfaces and various parameters. 
The official documentation\footnote{\url{https://www.mathworks.com/help/simulink/block-libraries.html}} provides specifications for inports, outports and parameters for each block type in natural language.

We formally re-describe the content of each block document as a \emph{feature model}~\cite{benavidesAutomated2010, johansenProperties2011} by identifying a set of \emph{features}; for example, parameter settings, enabled ports, and port data types are regarded as interdependent features.
An example (partial) specification for the \inlst{Product} block type, formatted in YAML style, is shown in Fig.~\ref{f:bspec}.
Specifications mainly consist of three sections, \inlst{Inputs}, \inlst{Output} and \inlst{Parameters}, each containing a list of mappings.
For blocks that take a variable number of inputs, we describe them separately for each number of inputs; here, we consider blocks with two inports named \inlst{Port_1} and \inlst{Port_2}.
Each mapping is named with a key \inlst{Name} (except for a single \inlst{Output} item), and also has a key \inlst{Domain} whose value is a set of data types (e.g. \inlst{Port_1}) or a list of values (e.g. \inlst{Inputs}).
We abbreviate the \Simulink data types with \inlst{T_float} (representing \inlst{double}, \inlst{float}, etc.), \inlst{T_int} (representing \inlst{int8}, \inlst{uint8}, \inlst{int32}, etc.) and \inlst{T_bool} (i.e. \inlst{boolean}).
Each input and output port has a key \inlst{Multiplicity} to specify whether possible signal values are scalar, vector, etc.
Items may have dependencies or additional constraints.
For example, the dependency of the domain of the outport being the same as inport \inlst{Port_1} is specified by referencing ``\inlst{<Port_1>}.''
The key \inlst{Constraint} specifies how its multiplicity is determined depending on the multiplicity of \inlst{Port_1} and \inlst{Port_2}.

Overall, we formalized 10 block types, three of which were specified in two ways for instances of unary and binary operations.
The first section of Tab.~\ref{t:result} summarizes the specified block type feature models.

\subsection{Test Suite Specifications}
\label{s:method:spec}

Test suites for $\ConfSL$ 
are specified by the following items.
\begin{itemize}
    \item \emph{Block type feature model}. 
    Each TC is an instance of a feature model,
    which is a minimal \Simulink model that mainly consists of a single block.
    \item \emph{Input signal specification}. 
    We consider fixed-length ($k$) input signals to the target block.
    We manually prepare a table that collects scalar values for each data type for the entire test process.
    %
    \item \emph{\Simulink as oracle}.
    The \Simulink model generated from a test vector is harnessed with an input signal builder;
    then, it is simulated with \Simulink to obtain the output signal, which is embedded in the verdict subsystem of the TC.
\end{itemize}

Test suites for $\ConfMCs$ are also specified by the following items, while uncertain aspects on signal values are additionally introduced. 
In the experiment, we assume a test suite once prepared for $\ConfSL$ and mutate the TCs so as to introduce uncertainties in the harnessed property monitors.
\begin{itemize}
    \item \emph{Block type feature model}. 
    \item \emph{Input signal specification}.
    For every step $l \in \{0,\ldots,k\!-\!1\}$ and for every $j$-th element of the input variable vector $\VIn = (i_1,\ldots,i_n)$, 
    the input signal builder for the first test specifies the value as $i_{j}(l) = \iota_{j}(l)$,
    where $\iota_{j}(l)$ is a prescribed constant.
    To make it uncertain, we either change it randomly to the constraint $i_{j}(l) \leq \iota_{j}(l)$, change it to $i_{j}(l) \geq \iota_{j}(l)$, or leave it unchanged.
    In the following, we denote the conjunction of equalities and inequalities for all elements by $\VIn^{[k]} \approx \iota^{[k]}$.
    \item \emph{\SmtMC as oracle}.
    Checking $\ConfMCs$ involves comparison of the results of \SmtMC and \SLDV, and we use the results from \SmtMC as a pseudo-oracle because it is more efficient and does not produce errors (as we will see in Sect.~\ref{s:xp}).
    Also for the comparison of output signals with expected values $\Int{M}_\mathrm{SL}(\iota^{[k]})$, we randomly make some element-wise comparisons into inequalities (the overall conjunction is denoted by $\VOut^{[k]} \approx \Int{M}_\mathrm{SL}(\iota^{[k]})$).
    We then verify the property $\varphi :\leftrightarrow \Box(\VIn^{[k]} \approx \iota^{[k]} \to \VOut^{[k]} \approx \Int{M}_\mathrm{SL}(\iota^{[k]}))$ with \SmtMC and obtain an expected value in $\{\True,\False\}$.
\end{itemize}

\IncMargin{1em}
\begin{algorithm}[t]
    \SetAlgoLined
    \SetKwInOut{Input}{Input}\SetKwInOut{Output}{Output}
    \Input{Specification $\mathit{spec}$, \# steps $k$, \\
    Flag $\mathit{doAbst}\!\in\!\{\True,\False\}$ \\
    indicating whether to check $\ConfMCs$ }
    \Output{Test statistic $\mathit{stat}$}
    \BlankLine
    $\mathit{TV} := \mathsf{runPICT}(\mathit{spec})$;\ $\mathit{stat} := \emptyset$\;
    \For{$t \in \mathit{TV}$}{
        $M := \mathsf{generateModel}(t)$\;
        $\Omic^{[k]} := \mathsf{runSimulink}(M,k)$\;
        $M := \mathsf{addVerdict}(M,\Omic^{[k]})$\;
        \textbf{if}\ {$\mathit{doAbst}$}\ \textbf{then}\ {$M := \mathsf{abstractHarness}(M)$;}\ \textbf{end}\\
        $\mathit{stat} := \mathsf{updateStat}(\mathit{stat}, \mathsf{runMC}(M))$\;
    }
    \Return{$\mathit{stat}$}\;
    \caption{A conformance testing procedure.}
    \label{a:testing}
\end{algorithm}

\subsection{Testing Procedure} 
\label{s:method:proc}

Alg.~\ref{a:testing} shows the procedure for each test.
When we test $\ConfSL$ of either \SmtMC or \SLDV, $\mathit{doAbst}$ is set to $\False$, and
when we test $\ConfMCs$ between \SmtMC and \SLDV, we set the flag to $\True$.
The algorithm measures the pass rate by counting the number of TCs that satisfy the condition~\eqref{eq:c1} or \eqref{eq:c2} in Def.~\ref{d:conf}.

First, it generates a set $\mathit{TV}$ of test vectors using a combinatorial testing tool.
In advance, we translate the block type feature model and the input signal specification into the format of the testing tool so that the set of generated vectors covers pairwise combinations of test factors.
%
Second, the algorithm generates a \Simulink model $M$ that instantiates $t \in \mathit{TV}$; it also embeds a concrete input signal specified by $t$ using \inlst{Constant} blocks.
Third, it runs \Simulink to obtain expected output signal values ($\Omic^{[k]}$) and constructs a verdict that compares output signals from the test block with the values in $\Omic^{[k]}$.
Finally, it runs the target MC tool to check $\ConfSL$ when $\mathit{doAbst}$ is $\False$; accordingly, the overall test result is updated.

When $\mathit{doAbst}$ is $\True$, 
the process checks $\ConfMCs$ for \SLDV under uncertainties; namely, 
we assume a set of input signals instead of a concrete one and test whether or not \SLDV can verify conditions on the output signals as \SmtMC does.
The $\mathsf{abstractHarness}$ procedure transforms a model $M$ for that purpose.
Given concrete input and output signals for $M$, it first introduces some uncertainties as explained in Sect.~\ref{s:method:spec} and encodes them in the harness of $M$.
It then verifies the consistency between the input and output constraints using \SmtMC and records the result in the harness as a verdict.

\subsection{Implementation}
\label{s:impl}


\begin{figure*}[t]
    \centering
    \includegraphics[width=\textwidth]{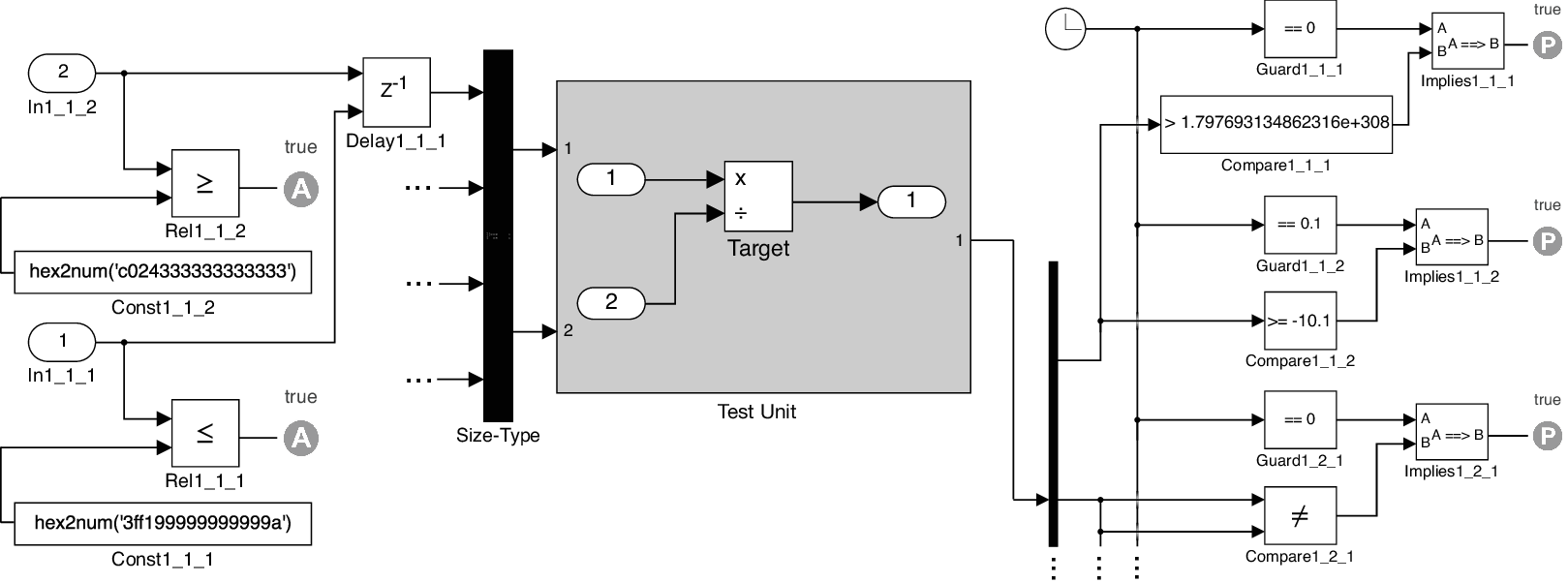} 
    \caption{Simulink harness using SLDV property blocks.}
    \label{f:harness}
\end{figure*}

\begin{figure*}[t]
    \begin{lstlisting}[basicstyle=\ttfamily\footnotesize,frame=single]
(define-fun trans ( (i#1#1 Float64) (i#1#2 Float64) (i#1#3 Float64) (i#2#1 Float64)
                    (o#1#1 Float64) (o#1#2 Float64) (o#1#3 Float64) ) Bool
  (and (= o#1#1 (fp.div RNE i#1#1 i#2#1))
       (= o#1#2 (fp.div RNE i#1#2 i#2#1)) (= o#1#3 (fp.div RNE i#1#3 i#2#1)) ))

(declare-const i#1#1@0 Float64) (declare-const i#1#2@0 Float64)
(declare-const i#1#3@0 Float64) (declare-const i#2#1@0 Float64)
(declare-const o#1#1@0 Float64) (declare-const o#1#2@0 Float64) (declare-const o#1#3@0 Float64)
(assert (fp.leq i#1#1@0 (fp #b0 #b01111111111 #b0001100110011001100110011001100110011001100110011010)))
(assert (= i#1#2@0 (_ +zero 11 53)))
(assert (fp.geq i#1#3@0 (fp #b1 #b11111111110 #b1111111111111111111111111111111111111111111111111111)))
(assert (= i#2#1@0 (_ +zero 11 53)))
(assert (trans i#1#1@0 i#1#2@0 i#1#3@0 i#2#1@0 o#1#1@0 o#1#2@0 o#1#3@0))
(check-sat-assuming ((not (= o#1#1@0 (_ +oo 11 53)))))
(check-sat-assuming ((not (= o#1#2@0 (_ NaN 11 53)))))
;; ... Other checks are omitted.
\end{lstlisting}
    \caption{SMT-LIB encoding of the bounded execution.}
    \label{f:smt:exec}
\end{figure*}

We have implemented the conformance testing method (Alg.~\ref{a:testing}) in the MATLAB scripting language. 
This implementation integrates \Simulink, \SLDV, and the \SmtMC tool, which is also implemented as a MATLAB script and interacts with the \Impl{Z3} SMT solver~\cite{demouraZ32008} (version~4.13.0 or 4.14.1).
We used \PICT~\cite{czerwonkaPairwise2006}
(commit \verb|7d2ed98|)\footnote{\url{https://github.com/microsoft/pict}} 
for combinatorial testing.

\vspace{.5em}
\subsubsection{Testing of \SLDV}
When the MC tool under test is \SLDV, our implementation generates a \Simulink harness model and feeds it to \SLDV.
An example harness model is shown in Fig.~\ref{f:harness}. 
The target block in the \inlst{Test Unit} subsystem is an instance of \inlst{Product} representing an element-wise division of type $\mathbb{F} \times \mathbb{F}^3 \to \mathbb{F}^3$ where $\mathbb{F}$ denotes the set of \verb|double| FP numbers.
The model is modified with $\mathsf{abstractHarness}$ to check $\ConfMCs$.
The fragment on the left of Fig.~\ref{f:harness} describes the following assumption on the input variable $i_1$ and the signals of length $2$.
\begin{equation*}
    i_{1}(0) \leq \Dbl(10.1) \land i_{1}(1) \geq \Dbl_\Max \land \ldots,
\end{equation*}
where $i_{j}(l)$ represents the $j$-th element of the value at step $l$.
In the initial step $0$, each inport is fed the input value to the target block for all steps, e.g., \inlst{In1_1_2} is initially fed $i_{1}(1)$ for the step~$1$; the values are then concatenated using delay blocks, e.g., \inlst{Delay1_1_1} receives a signal to be delayed and an initial value.
We have an inequality (or equality) for each signal value.
Blocks labeled as \verb|true| in the diagram are \SLDV assumption blocks, so the inequality expressions are interpreted as the conditions on the input signals by \SLDV.
Constant numerals are encoded exactly using the hexadecimal notation.
The \inlst{Size-Type} block combines the scalar signals into vector signals if necessary.
On the right side of Fig.~\ref{f:harness}, the model fragment specifies a verdict based on the following condition on the output variables $o_{1,1}$ and $o_{1,2}$.
\begin{multline*}
    o_{1,1}(0) = \Dbl(+\infty) \land o_{1,1}(1) \leq \Dbl(-10.1) \\
    \land o_{1,2}(0) = \Dbl(\Nan) \land \ldots
\end{multline*}
The model assumes that one step is $0.1$~seconds, and conditions are described using \SLDV implication blocks; for example, $o_{1,1}(0)$ is expected to be greater than the largest normal number by \inlst{Compare1_1_1}.
The constant numerals are again encoded exactly but rounded decimals are shown in the diagram.
To detect $\Nan$, it uses an inequality block (\inlst{Compare1_2_1}) because $\Nan = \Nan$ evaluates to $\False$.

\vspace{.5em}
\subsubsection{Testing of \SmtMC}
For \SmtMC, our implementation modifies its process and adds test verdict formulas to the encoded formula of the target \Simulink model.
Fig.~\ref{f:smt:exec} shows an SMT-LIB description for the example TC in Fig.~\ref{f:harness} associated with verdict formulas.
The upper part defines predicate \inlst{trans} representing the transition predicate of the \inlst{Product} block.
Next, the first transition is described by asserting the predicate with the constant symbols e.g. \inlst{i#1#1@0};
each input value is constrained by comparing it with a constant FP number.
In the last part, the constraints are checked for each output symbol.
To validate $\psi :\leftrightarrow o_{1,1}(0) = \Dbl(+\infty)$, we check the unsatisfiability by assuming $\neg\psi$.

%

\section{Experimental Results}
\label{s:xp}

We conducted experiments to answer the following research questions.

\vspace{.5em}

\begin{mdframed}
    \textbf{RQ1}: What are the pass rates for $\ConfSL$ and $\ConfMCs$ achieved by \SmtMC and 
    \SLDV?
    \textbf{RQ2}: If the test results are not conformant, what are the reasons?
    \textbf{RQ3}: What is the difference in test execution time between tools?
\end{mdframed}

Experiments were conducted in Docker containers (under the limitation of using up to 4~cores and 8GB RAM),
running on a 
4.5GHz AMD Ryzen~9 7950X 
processor. 
We ran \SmtMC and \SLDV on MATLAB~R2025b.
\SLDV was also run on R2020b for comparison; we refer to the R2020b and R2025b versions as \SLDVOld and \SLDVNew, respectively.
\SLDV was basically set to default settings, but some parameters were set as follows:
\inlst{MaxTestCaseSteps} to $10000$;
\inlst{MaxViolationSteps} to $1$ or $3$;
\inlst{DetectInfNaN} enabled; 
\inlst{DetectDivisionByZero} disabled.
The time limit for each TC was set to $300$s for both \SmtMC and \SLDV.

\begin{table}[!t]
    \centering
    \footnotesize
    \caption{Data types and the sampled values.} \label{t:data}
    \setlength{\tabcolsep}{.5em}
    \begin{tabular}{@{}llrl@{}}
        \toprule
        Category & Simulink type & \# & Values \\
        \midrule
        \lstinline|T_float|  & \verb|double| & 10 & $\mp\texttt{realmax('double')}$, $0$, $1$, etc. \\
        \lstinline|T_int|    & \verb|int8|  & 9 & $-127$, $-10$, $-1$, $0$, $1$, $2$, $19$, $127$, $128$ \\
        \lstinline|T_uint|   & \verb|uint8| & 6 & $0$, $1$, $2$, $23$, $254$, $255$ \\
        \lstinline|T_bool|   & \verb|boolean| & 2 & $\True$ and $\False$ \\
        \lstinline|T_string| & \verb|string|  & 2 & ``'' and ``123'' \\
        \bottomrule
    \end{tabular}
\end{table}

%
The input signal specifications were prepared as follows.
Note that we deliberately prepared a small specification to allow for manual analysis of failure cases in the experiment.
Testing using expanded block specifications and enhanced test signals remains a future challenge.
First, we set the signal length to 3 for stateful blocks (\inlst{Delay}, \inlst{UnitDelay} and \inlst{DiscreteIntegrator}) and 1 for other stateless blocks
(length~1 is sufficient to verify the correctness of the interpretation of such blocks).
Second, the test vectors enumerated by \PICT included the \inlst{Domain} and \inlst{Multiplicity} attributes.
For each data type category in the specification, we map a \Simulink data type as shown in Tab.~\ref{t:data}.
Between 2 and 10~sample values were prepared for each category.
For example, the list of values for type \inlst{double} contained some integers around $0$, maximal normal FP numbers, and other normal FP numbers.
The \inlst{Multiplicity} attribute specified whether the values are scalar or vector (we do not handle other composite signals e.g. matrices and buses, leaving them for future work).

Experimental results are shown in Tab.~\ref{t:result}.
Each row of the table corresponds to a block type (we abbreviate \inlst{DiscreteIntegrator} as \inlst{DiscInteg}).
The second section shows basic statistical data: The numbers of inports, block parameters, and the test vectors generated by \PICT.
The following sections show the results for each tool:
The columns ``\#~$\ConfSL$'' and ``\#~$\ConfMCs$'' show the numbers of TCs (and the pass rates) that were confirmed to satisfy the condition~\eqref{eq:c1} or \eqref{eq:c2}.
The time taken to process the passed TCs is aggregated in the ``Time'' column.
Although we did not need to test $\ConfMCs$ by targeting \SmtMC, we show the time required to check the same abstracted test conditions in the column ``$\textrm{Time}^*$.''

The results for $\ConfMCs$ are presented in the sections for \SLDVOld and \SLDVNew; note that these results represent comparisons between \SmtMC and the two versions of \SLDV, respectively.
As analyzed below, some of the non-conforming TCs were due to \SLDVOld or \SLDVNew not terminating successfully. \SmtMC terminated normally in all TCs.

For both \SLDVOld and \SLDVNew and all TCs, with the exception of the tests for \inlst{DiscInteg} on \SLDVOld, the cause of non-conforming cases for relation~\eqref{eq:c1} was the failure of \emph{compatibility} checking, which verifies whether the target model is described with the constructs supported by the model analysis process of \SLDV.
Unsuccessful \inlst{DiscInteg} TCs on \SLDVOld were due to timeouts.
Tests for $\ConfMCs$ failed due to various causes.
The ``Breakdown'' column lists the causes in the form ``$i$ / $j$ / $k$,''
meaning that:
$i$~failures are due to time running out;
$j$~tests terminate with internal error reports;
$k$~failures are due to non-conforming results.

The TC Simulink models, the encoded SMT-LIB files, and the experimental data are made public at \\
\url{https://doi.org/10.5281/zenodo.19464651}.

%

%
\begin{sidewaystable}
\centering
\caption{Experimental results.} \label{t:result}
\xrtable
\end{sidewaystable}
%

\section{Discussion}
\label{s:discuss}

In this section, we consider the pass rate to be the degree of conformance and confirm it for each tool.
We classify the test results and analyze their causes. In particular, we report the result of manual analysis of all TCs that failed 
with \SLDV.
In addition, we compare the performance of the MC tools in terms of test execution time.

\subsection{Pass Rate of \SmtMC for $\ConfSL$ (RQ1)}

\begin{mdframed}[style=answer]
    \textbf{Answer to RQ1}:
    \SmtMC passed all TCs for all blocks.
\end{mdframed}

\vspace{-.25em}

We consider the implementation of its encoder to be part of the block specification, so this result was necessary.
We repeated the test and debug loop multiple times to improve the tool's conformance to \Simulink.
As explained in Sect.~\ref{s:simulink:prod}, we analyzed the detailed semantics of each block type and the details of each numerical operation, and implemented them in the encoding process from blocks into logic formulas.
All TCs that yielded results different from \SLDV were inspected, and we found a number of bugs for the earlier versions; in the end, no bugs were found for the latest version.

\subsection{Pass Rate of \SLDV for $\ConfSL$ (RQ1 and RQ2)}

\begin{mdframed}[style=answer]
    \textbf{Answer to RQ1}:
    The average overall pass rates for \SLDVOld and \SLDVNew were $94\%$ and $96\%$, respectively.
\end{mdframed}

\vspace{-.5em}

\begin{mdframed}[style=answer]
    \textbf{Answer to RQ2}:
    All TCs (but one) represented computations resulting in the infinity FP numbers, and \SLDV failed with a message ``Simulink Design Verifier does not support non-finite numbers.''
    The 1~TC that passed only with \SLDVNew was for \inlst{Product1}, and its expected output value was $\Nan$.
\end{mdframed}

We did not describe any non-finite numeric literals in any TC.
For example, for a TC describing the addition $o_1 := i_1 + i_2$, we only set finite numbers as $i_1$ and $i_2$ and if $+\infty$ was expected for $o_1$, we monitored it by checking $o_1 > \mathit{MV}$, where $\mathit{MV}$ is a maximum finite number.
The number of TCs that passed in both versions differed by $36$, mainly because some TCs for \inlst{DiscInteg} timed out.
By block type, there were non-conforming cases in \inlst{Sum}, \inlst{Gain} and \inlst{Product}.


\subsection{Pass Rate of $\ConfMCs(\SmtMC, \SLDV)$ (RQ1 and RQ2)}

\begin{mdframed}[style=answer]
    \textbf{Answer to RQ1}:
    The average pass rate when comparing \SmtMC with \SLDVOld or \SmtMC with \SLDVNew were $80\%$ and $90\%$, respectively.
\end{mdframed}

Both \SLDVOld and \SLDVNew achieved a pass-all result only for simple block types (i.e. \inlst{Constant}, \inlst{Sum1}, \inlst{Abs} and \inlst{Logic}).

\vspace{.5em}
\begin{mdframed}[style=answer]
    \textbf{Answer to RQ2}:
    Some of the non-conforming TCs were due to timeouts, errors, and ``undecided'' outputs by \SLDV as broken down in Tab.~\ref{t:result}.
    Others were due to the handling of infinity FP values.
\end{mdframed}


\noindent\emph{Conclusiveness}.
\SmtMC completed all TCs successfully within the time limit, although there were some discrepancies with the results from \SLDV in several cases.

\vspace{.5em}
\noindent\emph{Timeouts}.
The tests for \inlst{Product2}, \inlst{Delay}, \inlst{UnitDelay} and \inlst{DiscInteg} took a long time to run on average and a total of $178$ with \SLDVOld and $104$ with \SLDVNew resulted in timeouts.
The main reason for the three stateful blocks was probably that the number of steps was set to 3 instead of 1, which resulted in the tripled number of properties and a large search space.
Also, nonlinear multiplication terms require special handling in the MC process, so we consider that checking \inlst{Product} and \inlst{DiscInteg} blocks was more expensive than checking others.

\vspace{.5em}
\noindent\emph{Errors}.
In tests on several blocks, 
8~errors in total occurred with \SLDVOld.
The exact reason was unknown but \SLDV reported that they were ``internal errors'' and were annotated as ``\verb|data_graph_err_internal|.''
With \SLDVNew, this error did not occur, but whenever we ran more than about 100~TCs in a row, the program would crash in a way that was difficult to reproduce.

\vspace{.5em}
\noindent\emph{Non-conforming results}.
In other TCs, the results of checking the property blocks by \SLDV did not match the results with \SmtMC, or \SLDV outputted ``undecided.''

According to our analysis, the reason for the former case, i.e., when the results were conclusive but non-conforming, was that \SLDV ignored executions where the results became infinite FP numbers.
\SLDV concluded these TCs as valid although an execution that resulted in an infinity value was violating the property.
These cases were the most common causes of non-conforming cases for block types \inlst{Sum2}, \inlst{Product2}, \inlst{Switch}, \inlst{Delay} and \inlst{UnitDelay}. 
There were $57$ such TCs with \SLDVOld and $101$ with \SLDVNew;
many such TCs with \SLDVNew were inconclusive with \SLDVOld.
In addition, we found that TCs for other block types contained similar executions and MC was performed as expected.

Other non-conforming test results were caused by some properties being judged as ``undecided.''
From the output, these can be classified into one of the following.
\begin{itemize}
    \item ``\emph{Undecided due to division by zero}.'' It occurred when a TC for \inlst{Product} expressed such an execution.
    There were $41$ such TCs with \SLDVOld and $16$ with \SLDVNew.
    \item ``\emph{Undecided with counterexample}.'' 
    This happened in some TCs ($2$ with \SLDVOld and $6$ with \SLDVNew) for \inlst{Product2}.
    For example, it occurred for a TC describing an operation $o_1 := (1/i_1) \times i_2$
    where $i_1$ and $i_2$ are assumed to be the maximum finite FP number and $-1.1$, respectively, and $o_1$ is expected to be less than a positive normal FP number.
    The actual cause was unknown to us.
    %
    %
    \item ``\emph{Undecided}.''
    Most of the failed TCs for \inlst{Delay}, \inlst{UnitDelay} and \inlst{DiscInteg} with \SLDVOld outputted no additional information and the reasons were unclear.
    These TCs were conclusive with \SLDVNew but ignored the infinity cases.
\end{itemize}

\subsection{Comparison of Execution Times (RQ3)}

\begin{mdframed}[style=answer]
    \textbf{Answer to RQ3}:
    In the tests for $\ConfMCs$ other than \inlst{Product}, \SmtMC was $3$ to $22$ times faster than \SLDVOld and $6$ to $16$ times faster than \SLDVNew. 
    The main reason is probably the simplicity of the process. 
\end{mdframed}

\subsubsection{Comparison of test processes for $\ConfSL$ and $\ConfMCs$}
In the testing of \SmtMC, checking TCs with uncertainty generated for $\ConfMCs$ was slightly faster than checking concrete TCs for $\ConfSL$ for all block types except \inlst{Product}. 
In contrast, TCs of \inlst{Product} for $\ConfMCs$ took a considerable amount of time; we believe this is due to the large number of operation patterns resulting from overflow etc.

\SLDV took slightly to $57$~times longer to process abstracted TCs for $\ConfMCs$ 
than TCs for $\ConfSL$ 
(per TC, \SLDVOld/\SLDVNew was $6.3$/$2.6$~times longer on average).
\SLDV seemed to have difficulty handling uncertainty.

\vspace{.5em}
\subsubsection{Comparison between \SLDVOld and \SLDVNew}
\SLDVNew resulted in higher pass rates because it could handle more cases within the time limit without outputting ``undecided'' compared to \SLDVOld.

In the test process for $\ConfSL$, \SLDVOld had a shorter execution time than \SLDVNew. 
On average, it was $1.7$ times faster.
In the test of $\ConfMCs$, \SLDVOld was faster for $10$~block types and slower for
\inlst{Switch}, \inlst{UnitDelay} and \inlst{DiscInteg}.
With \SLDVNew, the number of tests that ran out of time decreased except for \inlst{DiscInteg}, and the tests of \inlst{UnitDelay} and \inlst{DiscInteg}, which were relatively resource-intensive, were processed about three times faster on average.
This suggests that the MC algorithms differ between the versions.

\section{Threats to Validity}
\label{s:validity}

We point out two threats to internal validity (I-1 and I-2) and a threat to external validity (E-1). 

\vspace{.5em}
\emph{I-1)~Diversity and completeness of TCs}.
The number of prepared TCs can be insufficient because the factors in the block type specifications and the samples for the input signals are limited.
Therefore, it is possible that we have overlooked some bugs or non-conforming processes in the model checkers.
Considering that we manually analyzed the causes of the failed TCs, we believe that we conducted a sufficient number of them.
We also consider that the number of TCs was sufficient to improve the \SmtMC encoder through debugging;
in the end, it appears that the encoder implements generalized algorithms for each block type.
Nevertheless, further testing is a future task.

\vspace{.5em}
\emph{I-2)~Adequacy of the testing approach}.
Another threat is that our approach may not be adequate for testing the quality of MC tools.
Other approaches include testing using artificial evaluation kits, as well as fuzzing and mutation testing based on practical \Simulink models, as described in Sect.~\ref{s:related}.
In comparison, we apply automated testing methods to the \Simulink toolchain with minimal manual intervention within a basic scope, and we consider this engineering effort to be reasonable.
There are threats that testing of key features may be insufficient, or that bugs which only appear within large models may go undetected.
There are threats regarding the test oracles.
The \Simulink simulator used as a basis of $\ConfSL$ might have bugs, compromising the consistency of the specifications we have formalized.
\SmtMC used for $\ConfMCs$ was validated only by the first test with a limited number of TCs, so its reliability might be doubtful;
however, it should be noted that \SmtMC is used as a pseudo-oracle in the process, and the two tools are being compared equally.

\vspace{.5em}
\emph{E-1)~Formalization of basic \Simulink block types}.

Inappropriate choice of the domain of TC generation can be a threat to external validity.
There is concern as to whether the practically developed models align with the set of models we have generated in this work.
We focused on the basic block types, but other block types or combinations of several blocks can be more important factors for checking the correctness of the model checkers.
%
As a first step from the simplest form, we dealt with the 10~basic block types.

\section{Related Work}
\label{s:related}

\subsection{Formalization of \Simulink}
\label{s:related:formal}

Tripakis et al.~\cite{tripakisTranslating2005} and Bourbouh et al.~\cite{bourbouhCoCoSim2020} propose to convert a subset of the \Simulink language to Lustre, a synchronous programming language.
Bouissou and Chapoutot~\cite{bouissouoperational2012} formalize the integration process for discrete and continuous models and basic computation by some blocks.
Zhan et al.~\cite{zhanFormal2017} propose a formalization based on process algebra (CSP).
Bourke et al.~\cite{bourkeMechanized2020} formalize a \Simulink-like language in Coq and provide a validated code generation method.
Some~\cite{tripakisTranslating2005,bourbouhCoCoSim2020,bourkeMechanized2020} translate \Simulink models into transition systems, which is similar to our method, but differ in that they do not consider conforming them to bit-precise numerical computation implemented in \Simulink.
We are also unique in that we focus on the diversity of block instances and attempt to cover all parameter settings.

\subsection{Qualification of \Simulink, \SLDV, etc.}
\label{s:related:qual}

As the standards e.g. ISO~26262 and DO-178C demand, confidence in the toolchain used and qualification of each tool are important in the model-based development of safety-critical systems~\cite{wagnerFormal2017,ibrahimState2021}.
There is research on testing and verifying the practicality of the tools for modeling and MC, respectively;
our work belongs to the latter.

Chowdhury et al.~\cite{chowdhuryCyFuzz2017,chowdhuryAutomatically2018,chowdhurySLEMI2020} propose a series of testing methods and tools, {CyFuzz}, SLforge and SLEMI, for finding bugs in \Simulink.
Their methods are based on random testing and a collection of models, and some methods also employ techniques such as EMI-based mutation testing and differential testing.
SLforge and SLEMI have reported finding 8 and 9~new bugs, respectively.
Jiang et al.~\cite{jiangPartition2023} propose a method for finding bugs in \Simulink code generation by 
performing differential testing on multiple code generation processes; as a result, they have reported 11~new bugs.
In this paper, we do not target \Simulink but use it as an oracle as is, and conduct combinatorial tests to verify the conformance of model checkers.

The standard DO-330 recommends certifying the tool when the correctness of the tool output alone is not clear; also, DO-333 describes how to use formal methods in development~\cite{wagnerFormal2017,ibrahimState2021}. 
The result of a SWOT analysis~\cite{gleirscherQualification2023} organizes the perspectives for qualification of formal method tools used in CPS development.
In response, certification kits for MC tools, including ones from MathWorks,\footnote{\url{https://www.mathworks.com/products/iec-61508.html}} have been developed.
For example, the qualification package of the Kind2 model checker includes TCs that take into account the features of the modeling language~\cite{wagnerFormal2017}.
The test suite generated by our proposed method addresses the basic features of the \Simulink language and can be regarded as a kind of qualification kit, but the features it covers are limited; on the other hand, it is more precise than other kits in considering the numerical semantics of each block type.
%

There are several case studies that evaluate the usefulness and shortcomings of \SLDV functions and performance.
Nellen et al.~\cite{nellenFormal2018} have attempted verification using \SLDV against an in-vehicle system and pointed out gaps between industry needs and functions; similar to this work, they found that checking several properties took a long time and obtained many inconclusive results; they also obtained one spurious counterexample and pointed out reliability issues in numerical analysis.
Nejati et al.~\cite{nejatiEvaluating2019} have collected 10~industrial \Simulink models and analyzed the verification results of applying a test method and an SMT-based MC tool; as a result, they pointed out that, unlike testing, MC guarantees exhaustiveness but has issues with scalability.
Murray et al.~\cite{murraySafety2020} have compared the MC functions of \SLDV and RoboTool;
they also faced the analysis time issues and needed to simplify their models.
In this paper, we deal with the artificial \Simulink models that are prepared in a minimalist manner and do not evaluate scalability with respect to model size. 
However, we have confirmed that the execution time of \SLDV increases when properties are made uncertain. 
We have also conducted precise tests on numerical computation, which differs from other case studies.

\subsection{Bit-Precise Verification Methods}
\label{s:related:prec}

Numerical systems have been verified based on formalized FP numbers~\cite{harrisonFloatingPoint2006,boldoFlocq2011,brilloutMixed2009,brainAutomatable2015} and fixed-point numbers~\cite{baranowskiSMT2020,devadzeFormal2023}.
To our knowledge, only \SmtMC~\cite{ishiiSMTBased2022}, tested in this paper, verifies \Simulink models in a bit-precise way;
its analysis is based on a bit-blasting FP solver~\cite{brilloutMixed2009,brainAutomatable2015} and integer arithmetic operators implemented as in \cite{baranowskiSMT2020} based on the bit-vector theory.
The verification with our testing method is incomplete, so future tasks include verifying the numerical operation of the block types more exhaustively and formally, and generating provable certificates from the MC process.

\section{Conclusion}

We have described an empirical evaluation targeting two \Simulink model checkers, \SmtMC and \SLDV.
Our evaluation method is based on automatically measuring pass rates through block instantiation and checking the conformance conditions for $\ConfSL$ and $\ConfMCs$.
As deliverables, we developed block-type feature models, a modified bit-precise encoder of \SmtMC, and the test suite.
As a result of the evaluation, \SmtMC passed all TCs and confirmed the conformance to \Simulink within the scope of the specification, albeit incompletely.
We found some issues with the standard tool \SLDV.
In the testing for $\ConfSL$, the pass rate was $94$--$96\%$, due to the non-conforming TCs in the handling of infinity FP numbers.
For $\ConfMCs$, certain numbers of TCs were inconclusive ($186$ for R2020b and $104$ for R2025b), causing the pass rate to drop by about $9\%$ and $5\%$, respectively.
Overall, as the version number increased, the conformance rate between tools improved, and we identified the causes of failed TCs. 
As a result, we were able to obtain the implementation of \SmtMC that formally describes the language specification.
Currently limited to a subset of blocks and their features, we were able to verify whether the tools replicate the \Simulink simulator with bit-level precision.
We expect that this approach will improve reliability in CPS engineering.

As future work, more comprehensive specification descriptions and testing are required.
Specifically, we need to handle additional block types, generalize each block type specification, and improve the encoder. In addition, testing requires consideration of a larger number of sampled signals, combinations of several block instances, and verification of more generic properties.
This will provide a test suite that can be used as a qualification kit for the toolchain.

\section*{Acknowledgment}

The initial development of the testing tool was supported by KSE Software Solutions JSC.
This work was supported by JST, CREST JPMJCR23M1, and JSPS, KAKENHI 23K11969.

\bibliographystyle{IEEEtran}
\bibliography{Enc-Test}

\appendix

\section*{SMT-LIB Encoding of Example Simulink Models}
\label{s:smc:encoding}

This section introduces several examples of logical formulas in SMT-LIB generated by the \SmtMC tool from \Simulink models.

Fig.~\ref{f:smt:ex} shows an SMT-LIB description that encodes the \Simulink model in Fig.~\ref{f:ex}.
It first defines the predicates \inlst{init} and \inlst{trans} which represent the initial condition $\Init$ and the transition relation $\Trans$ of the underlying transition system (Def.~\ref{d:ts}), respectively.
Next, it describes a bounded execution (Def.~\ref{d:exec}) consisting of an initialization and two transitions by instantiating the predicates.
The default encoder of \SmtMC approximates \inlst{double} values as mathematical real numbers (of sort \inlst{Real}). In this example, all signal values are assumed to be of the \inlst{double} type and encoded accordingly.

Fig.~\ref{f:smt:prod1:double} shows a definition of the transition relation for the \inlst{Product} block instance in Fig.~\ref{f:ex:prod1} with each port set to type \inlst{double} (we do not use the \inlst{init} predicate for stateless blocks).
This example uses the exact encoder, which is an optional feature and which we have modified.
The exact encoder describes FP arithmetic using the \inlst{FloatingPoint} theory in SMT-LIB.
Sort \inlst{Float64} (abbreviation of \inlst{(_ FloatingPoint 11 53)}) represents 64-bit FP numbers and the \inlst{fp.mul} term represents the FP multiplication with the rounding mode RNE (round nearest ties to even).

Fig.~\ref{f:smt:prod1:int8} shows another definition of \inlst{trans} for the \inlst{Product} block instance in Fig.~\ref{f:ex:prod1} with each port typed as \inlst{int8}.
Since SMT-LIB does not include the theory of machine integers, we need to define our own operators.
We represent the value as a size-8 bit vector (of sort \inlst{(_ BitVec 8)}).
To replicate the behaviors described in Sect.~\ref{s:simulink:prod}, we carefully defined operators to satisfy $\ConfSL$. For example, we addressed settings for rounding or saturation (\inlst{SaturateOnIntegerOverflow}), and rounding direction (\inlst{RndMeth}).
First, we encode in a way that the intermediate results are maintained with high precision by converting the argument bit vectors to FP numbers.
To this end, we implement the multiplication function, \inlst{fp.mul_int8}, using \inlst{Float64} values (Line~28);
another branch handles overflow cases with the \inlst{int8.mul} operator (Lines~29--30), which may result in special values \inlst{int8.max_v} and \inlst{int8.min_v} (when the \verb|SaturateOnIntegerOverflow| parameter is on).
The first two branches of \inlst{fp.mul_int8} in Lines~20--23 bypass a special value when it is input.
Second, the auxiliary function, \inlst{int8.mul}, performs bit-vector operations with double-sized vectors, following the method in \cite{baranowskiSMT2020}.
Finally, the operator \inlst{int8.from_dbl} translates the internal \inlst{Float64} values back to machine integers.
Based on the block's parameter settings, it converts the maximum values, special values, and normal results, appropriately.

\begin{figure}[t]
    \begin{lstlisting}[basicstyle=\ttfamily\scriptsize,frame=single]
;; Initial condition.
(define-fun init ( (s13%1%1@0 Real) ) Bool
  (= s13%1%1@0 0) )

;; Transition relation.
(define-fun trans ( (s13%1%1@0 Real) (s13%1%1@1 Real) (i10%1%1 Real) (o10%1%1 Real) ) Bool
  (let ((lv15%1%1 (+ s13%1%1@0 (* 0.1 i10%1%1))))
  (and (= o10%1%1 s13%1%1@0) (= s13%1%1@1 (ite (> lv15%1%1 0) (- 1) lv15%1%1))) ) )

;; Execution trace.
(declare-const s13%1%1@i Real)
(assert (init s13%1%1@i))

(declare-const s13%1%1@0 Real)
(declare-const i10%1%1@0 Real)
(declare-const o10%1%1@0 Real)
(assert (trans s13%1%1@i s13%1%1@0 i10%1%1@0 o10%1%1@0))

(declare-const s13%1%1@1 Real)
(declare-const i10%1%1@1 Real)
(declare-const o10%1%1@1 Real)
(assert (trans s13%1%1@0 s13%1%1@1 i10%1%1@1 o10%1%1@1))
\end{lstlisting}
    \caption{SMT-LIB encoding of Ex.~\ref{f:ex} and its bounded execution.}
    \label{f:smt:ex}
\end{figure}

\begin{figure}[t]
    \begin{lstlisting}[basicstyle=\ttfamily\scriptsize,frame=single]
;; Transition relation.
(define-fun trans ( (i15%1%1 Float64)	(i15%2%1 Float64) (o15%1%1 Float64) ) Bool
  (= o15%1%1 (fp.mul RNE i15%1%1 i15%2%1) ))
\end{lstlisting}
    \caption{SMT-LIB encoding of Ex.~\ref{f:ex:prod1} typed as a \texttt{double} operator.}
    \label{f:smt:prod1:double}
\end{figure}

\begin{figure*}[t]
    \begin{lstlisting}[basicstyle=\ttfamily\scriptsize,frame=single]
;; Constants used by the int8 operators.
(define-const int8.max_v      (_ BitVec 8)  #b01111111)
(define-const int8.e_max_v_p1 (_ BitVec 9)  #b010000001)
(define-const int8.w_max_v    (_ BitVec 16) #b0000000001111111)
(define-const int8.min_v      (_ BitVec 8)  #b10000000)
(define-const int8.e_min_v_p1 (_ BitVec 9)  #b100000000)
(define-const int8.w_min_v    (_ BitVec 16) #b1111111110000000)

;; User-defined multiplication function for int8.
(define-fun int8.mul ((?saturate Bool) (x (_ BitVec 8)) (y (_ BitVec 8))) (_ BitVec 8)
  (let ((wx ((_ sign_extend 8) x)) (wy ((_ sign_extend 8) y)))
  (let ((doubled (bvmul wx wy)))
  (let ((wrapped ((_ extract 7 0) doubled)))
  (let ((saturated (ite (bvsgt doubled int8.w_max_v) int8.max_v
                   (ite (bvslt doubled int8.w_min_v) int8.min_v wrapped) ) ))
    (ite ?saturate saturated wrapped) )))) )

;; Double-typed wrapper for int8.mul.
(define-fun fp.mul_int8 ((?saturate Bool) (x Float64) (y Float64)) Float64
  (ite (= x ((_ to_fp 11 53) RNE int8.e_max_v_p1))
    x
    (ite (= x ((_ to_fp 11 53) RNE int8.e_min_v_p1))
      x
      (ite (or (fp.lt (fp.abs x) 
    (fp #b0 #b01111111111 #b0000000000000000000000000000000000000000000000000000))
               (fp.lt (fp.abs y) 
    (fp #b0 #b01111111111 #b0000000000000000000000000000000000000000000000000000)))
        (fp.mul RNE x y)
        ((_ to_fp 11 53) RNE (int8.mul ?saturate
          ((_ fp.to_sbv 8) RNE x) ((_ fp.to_sbv 8) RNE y))) ))) )

;; Casting function.
(define-fun int8.from_dbl ((?saturate Bool) (m RoundingMode) (x Float64)) (_ BitVec 8)
  (ite (= x ((_ to_fp 11 53) RNE int8.e_max_v_p1)) int8.max_v
    (ite (= x ((_ to_fp 11 53) RNE int8.e_min_v_p1)) int8.min_v
      (ite (fp.geq (fp.abs x)
        (fp #b0 #b11111111110 #b1111111111111111111111111111111111111111111111111111))
        (ite ?saturate int8.max_v #b00000000)
        (let ((doubled ((_ fp.to_sbv 16) m x)))
        (ite (and ?saturate (bvsge doubled int8.w_max_v)) int8.max_v
          (ite (and ?saturate (bvsle doubled int8.w_min_v)) int8.min_v
            ((_ extract 7 0) doubled) )))))) )

;; Transition relation.
(define-fun trans ( (i15%1%1 (_ BitVec 8)) (i15%2%1 (_ BitVec 8)) (o15%1%1 (_ BitVec 8)) ) Bool
  (= o15%1%1 (int8.from_dbl false RTN 
    (fp.mul_int8 false ((_ to_fp 11 53) RNE i15%1%1) ((_ to_fp 11 53) RNE i15%2%1))) ) )
\end{lstlisting}
    \caption{SMT-LIB encoding of Ex.~\ref{f:ex:prod1} typed as an \texttt{int8} operator.}
    \label{f:smt:prod1:int8}
\end{figure*}

\balance

\end{document}